\documentclass{iau-JDSS}
\usepackage{graphicx}

\title[ALMA] 
{ALMA}

\author[Turner \& Wootten]   
{Jean L. Turner$^1$  \and Alwyn Wootten$^2$\thanks{Present address: 
ALMA, 40 El Golf, Piso 18, Las Condes, Santiago, Chile.}
   }

\affiliation{$^1$Department of Physics and Astronomy, UCLA,
Los Angeles, CA 90095-1547 USA \break email: turner@astro.ucla.edu\\[\affilskip]
$^2$NRAO, 520 Edgemont Road, Charlottesville, VA 22903 USA\break email: awootten@nrao.edu}

\pubyear{2006}
\volume{Volume 14}  
\pagerange{119--126}
\date{?? and in revised form ??}
\jname{Highlights of Astronomy, Volume 14}
\editors{K.A. van der Hucht, ed.}
\begin{document}

\maketitle

\begin{abstract}
The Atacama Large Millimeter/Submillimeter Array (ALMA) is 
an international effort to construct an instrument 
capable of matching the exquisite imaging properties of optical space 
telescopes at millimeter and submillimeter wavelengths. ALMA science will transform our vision of the cold, dusty, and
gaseous universe, from extrasolar planets to the youngest galaxies.
\end{abstract}

\firstsection 
\section{Introduction to ALMA}

The
Atacama Large Millimeter/Submillimeter Array, ALMA, is the largest new facility for
long wavelength astronomy. It will be a large array of 
telescopes on a high, dry plain in the Atacama desert of northern Chile.
ALMA is a joint venture of the European Southern Observatory (ESO) and Spain, the National Radio Astronomy Observatory (NRAO) of Associated Universities Inc., funded by the U.S. National Science Foundation with support from the National Research Council of Canada, 
and the National Institutes of Natural Sciences of Japan with the Academia Sinica of Taiwan, in
cooperation with the Republic of Chile. The total projected cost of ALMA, based on the rebaseline of
the project completed in late 2005, is $\sim$1 billion current euros.  ALMA will consist of up to 64
12-m antennas, an additional 12 7-m antennas for mapping extended emission (the
Atacama Compact Array),  and 4 enhanced 12-m antennas capable of total power observing. 
ALMA will ultimately cover all atmospheric windows at wavelengths from 3mm to 350$\mu$m. 

ALMA is currently under construction at Llano de Chajnantor, near the historic town of San Pedro de Atacama. It shares the site with several existing and future millimeter and submillimeter facilities.
Construction highlights of the past year include the completion of
a road from the Operations Support Facility
at 2900m to the telescope site that wide enough to transport the antennas. The Array Operations
Site Technical Building is nearly complete at the 5000 meter level (16,570 ft), and is the largest steel-frame building in the world at this altitude. The ALMA Test Facility, an interferometer composed
of two antenna prototypes, is being tested at the VLA site. Antenna contracts have been awarded by the
partners, and the first 12-m antenna is due in Chile in 2007. 

\section{ALMA Science}

ALMA will be a mighty leap forward in capability, with
an order of magnitude gain in sensitivity over current 
millimeter arrays in the northern hemisphere, and the potential for
 two orders of magnitude gain in imaging resolution and fidelity,
at a high, dry site suited to submillimeter observing.
The ALMA Level-One design goals exemplify the types of projects that will drive this
transformational instrument.  These goals are: 1) to detect CO in an L* galaxy at z=3; 2) 
to image molecular
lines in a protoplanetary disk with a resolution of 1 AU out to a distance of 150 pc; and 3) to
obtain 
high fidelity imaging at 0$^{\prime\prime}$.1 to match HST, JWST, or AO imaging.  
ALMA will allow us to
study the gas in ``normal" galaxies at redshifts at which galaxies should be significantly 
different from the present, in the ``Gas Ages."  Closer to home,
ALMA will enable study of the birth of planetary systems, allowing imaging of
 the nearest protoplanetary disks with 1 AU spatial  resolution
and sufficient spectral resolution to resolve motions on planet scales within a gaseous protoplanetary
disk.

The science capabilities of ALMA span all fields of astronomy.

\noindent {\bf $\bullet$ ALMA and planet formation.} At its highest resolution of 5mas, ALMA will easily detect
thermal dust continuum emission from protoplanetary or debris disks out to the nearest star clusters at 
distances of 150 pc. It will be able to image the thermal emission from very young
Jupiter-mass protoplanets out to 100 pc. 

\noindent {\bf $\bullet$ ALMA and molecular gas.}  ALMA will image emission from lines of the 
over 140 molecules currently detected in space, a list that includes organic molecules of biological
interest, including amino acids. In addition to astrochemistry, studies of molecular abundances
and excitation in interstellar clouds will provide diagnostics for interstellar physics, from shocks
to gas dynamics, allowing the study of star formation and the feedback of star formation on molecular clouds.

\noindent {\bf $\bullet$ ALMA and nearby galaxies.} ALMA will image molecular clouds in 
galaxies to give us a new view of galactic structure, galaxy interactions and galaxy mergers, and star
formation on galactic scales. ALMA will probe
the gas dynamics within a few parsecs of supermassive black holes in the centers
of galaxies. At its highest resolution,
ALMA can directly resolve structures of 50 AU ($\sim$600 Schwarzschild radii) in Sgr A*.

\noindent {\bf $\bullet$ ALMA and the solar system.} The Submillimeter Array in Hawaii recently 
detected
the thermal emission from Pluto and Charon: while Charon's surface is the expected
50~K for this distance from the sun, 
Pluto's temperature is a chilly 38~K; sublimation of surface ice is 
suspected. 
With its unprecedented spatial resolution
and sensitivity, ALMA will  be able to image surface features on the outer planets,
in addition to  asteroids, dwarf planets and icy Kuiper belt objects. ALMA
 will easily detect and perhaps even
resolve the most distant object known in the solar system, UB313. The composition, structure, and 
kinematics of planetary atmospheres will also be within ALMA's reach.

\noindent {\bf $\bullet$ ALMA and the distant universe.} ALMA will be a unique eye on 
 the early universe. Actively star-forming galaxies, such as Arp 220, can be detected easily in CO
to z=10. Images of gas and dust emission will allow us to study many properties of 
primeval galaxies, such as structure,  dynamical masses, and gas masses. While
 the Hubble Deep field contains many nearby galaxies and few distant ones, the ALMA
 Deep Field is the opposite: to a given brightness, simulations show that
 galaxies beyond z=1.5 are a hundred times more numerous than nearby galaxies.

\section{When can I  use ALMA?}

After the first antenna arrives in Chile in 2007, there will be an extended
period of system integration, verification, and commissioning. 
  ``Early Science," when ALMA is first opened to the community, is scheduled to begin in
   2010, when the array has 12--16 antennas.  ALMA reaches full capability, with
   all antennas and receiver bands, in 2012. 
    More ALMA information and an ALMA Cam are available at http://www.alma.cl.
   












\end{document}